\documentclass[a4paper,reqno,12pt,draft]{amsart}
\usepackage{amssymb,euscript,bbold}
\usepackage{array}
\renewcommand{\d}{\partial}
\newcommand{\half}{\tfrac12}
\newcommand{\reg}{\text{reg}}
\renewcommand{\gg}{\mathfrak g}

\newcommand{\sG}{\mathsf G}
\newcommand{\sT}{\mathsf T}

\newcommand{\D}{\mathsf{D}}

\newcommand{\eC}{\EuScript{C}}
\newcommand{\eD}{\EuScript{D}}

\newcommand{\eO}{\EuScript{O}}
\newcommand{\1}{\mathbb 1}

\newcommand{\R}{\mathbb R}
\newcommand{\Z}{\mathbb Z}
\newcommand{\J}{\mathbb J}

\newcommand{\AdS}{\text{AdS}}
\newcommand{\opp}{\mathrm{opp}}
\newcommand{\Aut}{\mathrm{Aut}}
\newcommand{\Inn}{\mathrm{Inn}}
\newcommand{\Out}{\mathrm{Out}}
\newcommand{\SL}{\mathrm{SL}}
\newcommand{\SU}{\mathrm{SU}}

\newcommand{\SO}{\mathrm{SO}}
\newcommand{\su}{\mathfrak{su}}
\renewcommand{\sl}{\mathfrak{sl}}

\DeclareMathOperator{\diag}{diag}
\DeclareMathOperator{\Ad}{Ad}

\DeclareMathOperator{\pr}{pr}
\newtheorem{thm}{Theorem}

\begin{document}

\title{$\D$-branes in $\AdS_3\times S^3\times S^3\times S^1$}  
\author[Figueroa-O'Farrill]{Jos\'e M Figueroa-O'Farrill}
\address{\begin{flushright}
    Department of Mathematics and Statistics\\
    University of Edinburgh\\
    King's Buildings (JCMB)\\
    Mayfield Road\\
    Edinburgh EH9 3JZ\\
    Scotland\\
  \end{flushright}}
\email{jmf@maths.ed.ac.uk}
\author[Stanciu]{Sonia Stanciu}
\address{\begin{flushright}
    Institute for Theoretical Physics\\
    Utrecht University\\
    Princetonplein 5\\ 
    3584 CC Utrecht\\
    The Netherlands\\
  \end{flushright}}
\email{s.stanciu@phys.uu.nl}
\thanks{Edinburgh MS-00-002, Imperial/TP/98-99/61, ITP-UU-00/01}
\begin{abstract}
  We analyse the possible $\D$-brane configurations in an $\AdS_3\times S^3
  \times S^3\times S^1$ background with a NS-NS B field, by using the
  boundary state formalism.  We study their geometry and we determine
  the fraction of spacetime supersymmetry preserved by these solutions.
\end{abstract}
\maketitle

\section{Introduction}

In \cite{Sads3} we have initiated a study of the possible $\D$-brane
configurations in exact superstring backgrounds based on AdS spaces
and characterised by a NS-NS antisymmetric field.  Here we will
continue this study by considering the case of the $\AdS_3\times
S^3\times S^3\times S^1$ background.  String propagation on this
background was studied in \cite{EFGT}.  This geometry appears as the
throat limit of two differently oriented coincident sets of fivebranes
intersecting in one direction, together with a set of infinitely
stretched strings \cite{CT,BPS,GMT}.

The paper is organised as follows.  In the next section we start with
a short summary describing the bosonic background in order to set the
notation and exhibit the conformal structure.  In Section 3 we
consider the boundary state formalism adapted to this particular
model; we write down the gluing conditions which preserve conformal
invariance and the affine symmetry of the underlying current algebra
and solve for them, thus determining two classes of bosonic
configurations.  In Section 4 we identify the $\D$-brane
configurations that each of the two classes of solutions give rise to.
The first class of solutions describe $\D$-brane configurations which
can be thought of as a straightforward generalisation of the D-type
configurations in $\AdS_3\times S^3\times T^4$ \cite{Sads3}.  By
contrast, the second class of solutions has no such analogue in the
$\AdS_3\times S^3\times T^4$ background.  Therefore we devote Section
5 to determining the twisted conjugacy classes of $\SU(2)\times\SU(2)$
and analysing their geometry and topology.  In Section 6 we extend our
analysis to the $N{=}1$ supersymmetric case and we find that all the
bosonic configurations determined previously can be made into $N{=}1$
configurations.  We then analyse, in Section 7, the fraction of
supersymmetry preserved by these configurations, and we find that they
preserve half of the spacetime supersymmetry.  We end, in Section 8,
with a summary of results.

\section{The bosonic $\AdS_3\times S^3\times S^3\times S^1$ background}

Bosonic string propagation on the background $\AdS_3\times S^3\times
S^3\times S^1$ can be described using a free compact boson for the
$S^1$ factor, and a WZW model with semisimple group $\SL(2,\R) \times
\SU(2) \times \SU(2)$ for the $\AdS_2 \times S^3 \times S^3$ part.

The theory of the compact free boson for the $S^1$ factor is described 
by the action
\begin{equation*}
  I_{S^1}[\varphi] = \int_\Sigma \d\varphi\bar\d\varphi~,
\end{equation*}
where $\Sigma$ is an orientable Riemann surface.  The operator product
algebra is standard:
\begin{equation}
  \d\varphi(z)\d\varphi(w) = \frac{1}{(z-w)^2} + \reg~,\label{eq:S1}
\end{equation}
with similar operator product expansions for the antiholomorphic
sector.

The WZW model describing the $\AdS^3 \times S^3 \times S^3$ part has
as target the semisimple group $\mathbf{G}=\mathbf{G_1}
\times\mathbf{G_2}\times\mathbf{G_3}$, with $\mathbf{G_1}=\SL(2,\R)$,
and $\mathbf{G_2}$ and $\mathbf{G_3}$ two copies of $\SU(2)$.  The
corresponding action will therefore be a sum of three terms
\begin{equation}
I = I_{\SL(2,\R)}[g_1] + I_{\SU(2)}[g_2] +
    I_{\SU(2)}[g_3]~,\label{eq:wzw} 
\end{equation}
where each term is of the standard form
\begin{equation*}
\frac{2}{k_i}I[g_i] = \int_{\Sigma} \langle g_i^{-1}\d g_i,
                      g_i^{-1}\bar\d 
                g_i\rangle + {\textstyle \frac{1}{6} }\int_B \langle
                g_i^{-1}dg_i, [g_i^{-1}dg_i,g_i^{-1}dg_i]\rangle~,
\end{equation*}
with $\d B = \Sigma$.  Each of the fields $g_i$ is a map from $\Sigma$
to the Lie group $\mathbf{G_i}$, $i=1,2,3$.  We denote by
$\gg=\gg_1\oplus\gg_2\oplus\gg_3$ the corresponding Lie algebra, where
$\gg_1=\mathfrak{sl}(2,\R)$ and $\gg_2,\gg_3=\mathfrak{su}(2)$.  For
these algebras we choose the following bases of generators: $\{T_a\}$
for $\gg_1$, $\{X_a\}$ for $\gg_2$ and $\{Y_a\}$ for $\gg_3$
satisfying
\begin{equation*}\label{eq:gg1}
[T_1, T_2] = T_3~,\qquad [T_2, T_3] =- T_1~,\qquad [T_3, T_1] =- T_2~,
\end{equation*}
and
\begin{equation*}\label{eq:gg2}
[X_1, X_2] = X_3~,\qquad [X_2, X_3] = X_1~,\qquad [X_3, X_1] = X_2~,
\end{equation*}
and similar Lie brackets for the $Y_a$'s.  We also need to specify an
invariant metric on $\gg$, which has a diagonal form
\begin{equation*}\label{eq:eta}
\eta = \begin{pmatrix}
           \eta_1 & 0 & 0\\
           0 & \eta_2 & 0\\
           0 & 0 & \eta_3
         \end{pmatrix}~,
\end{equation*}
with three components given by $(\eta_1)_{ab} = \diag(+,+,-)$ and
$(\eta_2)_{ab} = (\eta_3)_{ab} = \diag(+,+,+)$.

A group element $g=(g_1,g_2,g_3)$ in $\mathbf{G}$ can be parametrised
as follows:
\begin{equation}
g_1 = e^{\theta_2 T_2}e^{\theta_1 T_1}e^{\theta_3 T_3}~,\qquad 
g_2 = e^{\phi_2 X_2}e^{\phi_1 X_1}e^{\phi_3 X_3}~,\label{eq:parg}
\end{equation}
\begin{equation}\label{eq:parg'}
g_3 = e^{\sigma_2 Y_2}e^{\sigma_1 Y_1}e^{\sigma_3 Y_3}~,
\end{equation}
where $\theta_{\mu}$, $\phi_{\mu}$ and $\sigma_{\mu}$, $\mu=1,2,3$,
play the r\^ole of the spacetime fields.  In terms of them
\eqref{eq:wzw} becomes a sigma-model action, whose spacetime metric
and antisymmetric field can be straightforwardly obtained (the
expressions for $\SL(2,\R)$ and $\SU(2)$ were explicitly written down
in \cite{Sads3}).

This model possesses, as is well-known, an infinite-dimensional
symmetry group $\mathbf{G}(z)\times\mathbf{G}(\bar z)$ characterised
by the conserved currents $\J(z)=- \d g g^{-1}$ and $\bar\J(\bar z)=
g^{-1}\bar\d g$,
which underlies the exact conformal invariance of this background.
The conserved currents generate, at the quantum level, an affine Lie
algebra, $\widehat\gg_1\oplus\widehat\gg_2\oplus\widehat\gg_3$,
described by
\begin{equation}
\J_a (z)\J_b (w) = \frac{h_{ab}}{(z-w)^2} + \frac{{f_{ab}}^c 
                   \J_c (w)}{z-w} + \reg~,\label{eq:affg}
\end{equation}
where the indices run over the whole Lie algebra, $a,b=1,...,9$, and
the coefficients $h_{ab}$ define the symmetric bilinear form
\begin{equation}\label{eq:g}
h = \begin{pmatrix}
      k_1\eta_1 & 0 & 0\\
      0 & k_2\eta_2 & 0\\
      0 & 0 & k_3\eta_3
         \end{pmatrix}~.
\end{equation}
The parameters $k_i$ are related to the level $x_i$ of the
corresponding affine algebra by $k_i=x_i+g_i^*$, where $g_i^*$ is the
dual Coxeter number.

The CFT corresponding to this string background is then described by
the energy-momentum tensor
\begin{equation*}
\sT = \Omega^{ab}(\J_a\J_b) + (\d\varphi\d\varphi)~, 
\end{equation*}
where $\Omega^{ab}$ are the components of the inverse $\Omega^{-1}$ of
the invariant metric
\begin{equation*}\label{eq:Omega}
\Omega = \begin{pmatrix}
           \Omega_1 & 0 & 0\\
           0 & \Omega_2 & 0\\
           0 & 0 & \Omega_3
         \end{pmatrix}~,
\end{equation*}
with the components given by $\Omega_1=2(k_1+1)\eta_1$,
$\Omega_2=2(k_2-1)\eta_2$, and $\Omega_3=2(k_3-1)\eta_3$.  The central
charge of this CFT is given by
\begin{equation*}
c = \frac{3k_1}{k_1+1} + \frac{3k_2}{k_2-1} + \frac{3k_3}{k_3-1} + 1~.
\end{equation*}

\section{Boundary states}
\label{sec:boundary}

The strategy that we will follow in order to determine the $\D$-brane
configurations which can be consistently defined in type IIB string
theory on $\AdS_3\times S^3\times S^3\times S^1$ is similar to the one
used in \cite{Sads3} and further developed in \cite{SDnotes}.  We
consider a class of gluing conditions, which is defined in terms of a
Lie algebra automorphism, $R:\gg\to\gg$, which preserves the metric
$\eta$:
\begin{equation}
[R(Z_a),R(Z_b)] = R([Z_a,Z_b])~,\label{eq:Rhom}
\end{equation}
\begin{equation}
R^T \eta R = \eta~,\label{eq:Rg}
\end{equation}
where $\{Z_a\}$ is a given basis in $\gg$, in terms of which $R$ is
given by $R(Z_a)=Z_b{R^b}_a$.  The gluing conditions read
\begin{equation}
\J_a(z) - {R^b}_a {\bar\J}_b(\bar z) = 0~,\label{eq:Dbc} 
\end{equation}
and can be easily seen to preserve the current algebra of the bulk
theory.

These gluing conditions have to satisfy the basic consistency
requirement, which is conformal invariance.  In the bosonic case, this
comes down to imposing
\begin{equation*}
\sT(z) = \bar\sT(\bar z)~,\label{eq:ci}
\end{equation*}
at the boundary.  In this case, the requirement of conformal
invariance translates into the condition
\begin{equation}
R^T \Omega R = \Omega~.\label{eq:Rmet}
\end{equation}

As explained in \cite{FSNW}, $\D$-branes in a WZW model with group
$\mathbf{G}$ are classified by the group $\Out_o(\mathbf{G})$ of
metric-preserving outer automorphisms of $\mathbf{G}$, which is 
defined as the quotient $\Aut_o(\mathbf{G})/\Inn_o(\mathbf{G})$ of the
group of metric-preserving automorphisms by the invariant subgroup of
inner automorphisms.  For the case at hand, and ignoring the $S^1$
factor for which no automorphism is inner, $\Out_o(\mathbf{G}) \cong
\Z_2$, whence there are two distinct types of $\D$-branes on
$\SL(2,\R) \times \SU(2) \times \SU(2)$.  Let us see this.

Group automorphisms for simply connected groups come by exponentiating 
automorphisms of the Lie algebra.  In our case, the Lie algebra $\gg$
of $\mathbf{G}$ is a direct sum of three terms $\gg_1$, $\gg_2$ and
$\gg_3$, each of them being a three-dimensional simple Lie algebra.
Since $\mathfrak{sl}(2,\R)$ and $\mathfrak{su}(2)$ are non-isomorphic
simple Lie algebras, there is no nontrivial homomorphism between them.
We therefore deduce that the matrix of boundary conditions defined by
the automorphism $R:\gg_1\oplus\gg_2\oplus\gg_3 \to
\gg_1\oplus\gg_2\oplus \gg_3$, must take one of the following two
forms
\begin{equation}\label{eq:R}
R_I = \begin{pmatrix}
           R_{11} & 0 & 0\\
           0 & R_{22} & 0\\
           0 & 0 & R_{33}
         \end{pmatrix}~,\qquad
R_{II} = \begin{pmatrix}
           R_{11} & 0 & 0\\
           0 & 0 & R_{23}\\
           0 & R_{32} & 0
         \end{pmatrix}~,
\end{equation}
where $R_{ij}:\gg_j\to\gg_i$, for any $i,j=1,2,3$.  We thus have two
classes of solutions: the first, described by $R_I$, exists for any
values of the parameters $k_i$, whereas the second, given by $R_{II}$,
exists only for particular values of the parameters, such that
$k_2=k_3$.  Moreover, from \eqref{eq:Rmet} and \eqref{eq:Rg}, it
follows that $R_{11}$ belongs to $\mathrm{O}(2,1)$, and $R_{ij}$, with
$i,j=2,3$, belong to $\mathrm{O}(3)$.  On the other hand, from
\eqref{eq:Rhom} we deduce that $R_{ii}$ are Lie algebra automorphisms,
corresponding to $\mathfrak{sl}(2,\R)$ and $\mathfrak{su}(2)$, whereas
$R_{23}$ and $R_{32}$ are Lie algebra isomorphisms.  Explicitly, each
of these conditions translates into a condition on the corresponding
matrix, that reads
\begin{equation*}
\det(R_{ij}) = 1~,
\end{equation*}
which makes $R_{11}$ belong to $\SO(2,1)$, and $R_{ij}$, for
$i,j=2,3$, to $\SO(3)$.

Clearly, the main difference between these two classes of solutions is
that $R_I$ describe inner automorphisms, whereas $R_{II}$ does not.
More precisely, we have
\begin{equation*}
R_{II} = T R_I~,
\end{equation*}
where the matrix $T$ is given by
\begin{equation*}
T = \begin{pmatrix}
           \1 & 0 & 0\\
            0 & 0 & \1\\
            0 & \1 & 0
    \end{pmatrix}~,
\end{equation*}
with obvious notation.

These results can be summarised as follows.  We consider a set of
gluing conditions on the group manifold $\SL(2,\R)\times
\SU(2)\times\SU(2)$ which preserve conformal invariance and the
infinite-dimensional symmetry of the current algebra of the bulk
theory.  These gluing conditions are described in terms of
metric-preserving automorphisms of the Lie algebra 
$\sl(2,\R)\oplus\su(2)\oplus\su(2)$.  They admit two classes of 
solutions, characterised by different matrices of gluing conditions:
for generic values of the levels $k_i$, the solutions are parametrised
by elements of $\SO(2,1)\times \SO(3)\times\SO(3)$; additionally, for
the particular values of $k_i$ such that $k_2=k_3$ we have an extra
set of solutions, parametrised again by the elements of the group
$\SO(2,1)\times\SO(3)\times \SO(3)$.

\section{$\D$-brane solutions}

Let us begin with the $\D$-brane configurations produced by the inner
automorphisms $R_I$.  This case represents a straightforward
generalisation of the corresponding analysis performed for the
$\AdS_3\times S^3\times T^4$ background.  Indeed, since the matrix of
gluing conditions $R_I$ is block diagonal, the resulting $\D$-brane
configurations take a product form, $\eD_{\SL(2,\R)}\times
\eD_{\SU(2)}\times\eD_{\SU(2)}$, where $\eD_\mathbf{G}$ represent the
$\D$-brane configurations in the group manifold $\mathbf{G}$.  The
possible $\D$-brane configurations in $\SL(2,\R)$ and $\SU(2)$ have
been studied in detail in \cite{AS,Sads3}.  We therefore obtain in our
case that the $\D$-brane solution passing through a point $g$ in
$\mathbf{G}$ and being described by a set of gluing conditions defined
by the inner automorphism $R_I$ is characterised by a worldvolume
which lies along a product of conjugacy classes shifted by group
elements determined by $R_I$:
\begin{equation*}
\eC_{\SL(2,\R)}(g_1r_1^{-1})r_1 \times \eC_{\SU(2)}(g_2r_2^{-1})r_2
\times \eC_{\SU(2)}(g_3r_3^{-1})r_3~,
\end{equation*}
where $R_{11}=\Ad_{r_1}$, $R_{22}=\Ad_{r_2}$, $R_{33}=\Ad_{r_3}$, for
some $(r_1,r_2,r_3)$ in $\mathbf{G}$.  The conjugacy classes of
$\SU(2)$ are parametrised by $S^1/\Z_2$, which we can understand as
the interval $\theta\in[0,\pi]$.  The conjugacy classes corresponding
to $\theta=0,\pi$ are points, corresponding to the elements $\pm e$ in
the centre of $\SU(2)$, whereas the classes corresponding to
$\theta\in(0,\pi)$ are 2-spheres.  If we picture $\SU(2)$, which is
homeomorphic to the 3-sphere, as the one-point compactification of
$\R^3$ where the sphere at infinity is collapsed to a point, the
foliation of $\SU(2)$ by its conjugacy classes coincides with the
standard foliation of $\R^3$ by 2-spheres with two degenerate spheres
at the origin and at infinity.  For $\SL(2,\R)$, on the other hand, we
have three types of metrically nondegenerate\footnote{This condition
  is necessary for their interpretation as D-branes.} conjugacy
classes (for details, see \cite{Sads3}): two point-like ones,
corresponding to the two elements in the centre of $\SL(2,\R)$, a
family of two-dimensional classes with planar topology and a family of
two-dimensional classes with cylindrical topology.

We now turn to the $\D$-branes produced by outer automorphisms $R_{II}$.
Notice first of all that, since the matrix of gluing conditions
$R_{II}$ has a block diagonal form, the resulting $\D$-brane
configurations take, also in this case, a product form,
$\eD_{\SL(2,\R)}\times\eD_{\SU(2)\times\SU(2)}$, where
$\eD_{\SL(2,\R)}$ represent the same $\D$-brane configurations in the
group manifold $\SL(2,\R)$ which were discussed above.  It remains to
analyse the $\eD$-brane configurations on the product group
$\SU(2)\times\SU(2)$, corresponding to the gluing conditions
\begin{equation}\label{eq:gtilde}
-\d\tilde g\tilde g^{-1}=\tilde R_{II} (\tilde g^{-1}\bar\d\tilde g)~,
\end{equation}
where $\tilde g=(g_2,g_3)$, and the corresponding matrix of gluing
conditions
\begin{equation*}
\tilde R_{II} = \begin{pmatrix}
                  0 & R_{23}\\
                  R_{32} & 0
                \end{pmatrix}
\end{equation*}
can be easily shown to be of the form
\begin{equation*}
\tilde R_{II} = \tilde T \Ad_{\tilde r}~,\qquad\qquad 
\tilde T = \begin{pmatrix}
                  0 & \1\\
                  \1 & 0
                \end{pmatrix}~,
\end{equation*}
where $\tilde r=(r_2,r_3)$, while $R_{32}=\Ad_{r_2}$ and
$R_{23}=\Ad_{r_3}$.  This implies that it is sufficient to analyse the
$\D$-brane configurations produced by $\tilde R_{II}=\tilde T$ in
$\SU(2)\times\SU(2)$; any other $\tilde R_{II}$ will lead to
configurations which differ from these only by translations in the
group manifold.  Indeed, a set of gluing conditions described by a
generic $\tilde R_{II}$
\begin{equation*}
-\d\tilde g\tilde g^{-1}=\tilde T\cdot\Ad_{\tilde r} (\tilde
g^{-1}\bar\d\tilde g)~,
\end{equation*}
for some $\tilde r$ in $\SU(2)\times\SU(2)$, can be written as
\begin{equation*}
-\d h h^{-1}=\tilde T (h^{-1}\bar\d h)~,
\end{equation*}
with $h=\tilde g\tilde r^{-1}$.  According to the general theory
developed in \cite{SDnotes} (for a somewhat different approach, see
\cite{FFFS}) the $\D$-brane configurations produced by the gluing
conditions \eqref{eq:gtilde} with $\tilde R_{II}=\tilde T$ are nothing
but the twisted conjugacy classes determined by the outer automorphism
$\tilde T$, which we denote by $\eC^{\tilde
  T}_{\SU(2)\times\SU(2)}(\tilde g)$.  In summary, the $\D$-brane
configurations passing through a point $g$ in
$\SL(2,\R)\times\SU(2)\times\SU(2)$ and described by the matrix of
gluing conditions $R_{II}$ have worldvolumes which lie along a product
of twisted conjugacy classes
\begin{equation*}
\eC_{\SL(2,\R)}(g_1r_1^{-1})r_1 \times \eC^{\tilde
T}_{\SU(2)\times\SU(2)}(\tilde g{\tilde r}^{-1})\tilde r~.
\end{equation*}
Thus, in order to complete our analysis, we have to determine the
twisted conjugacy classes of $\SU(2)\times\SU(2)$.  Next section will
be devoted to solving this mathematical problem.  There we will show
that the twisted conjugacy classes of $\SU(2)\times\SU(2)$
corresponding to the outer automorphism $\tilde T$ are basically
determined by the (ordinary) conjugacy classes of $\SU(2)$.  Indeed,
if we denote by $m$ the group multiplication,
$m:\SU(2)\times\SU(2)\to\SU(2)$, which assigns to every element
$(g_2,g_3)$ in $\SU(2)\times\SU(2)$ an element in $\SU(2)$ given by
$m(g_2,g_3)=g_2 g_3$, then we have that
\begin{equation*}
\eC^{\tilde T}_{\SU(2)\times\SU(2)}(g_2,g_3) = 
                               m^{-1} \eC_{\SU(2)}(g_2 g_3)~.
\end{equation*}
Using this result we will see that the group $\SU(2)\times\SU(2)$ has
two types of conjugacy classes: two three-dimensional ones,
diffeomorphic to $S^3$, and a family of five-dimensional classes,
diffeomorphic to $S^2\times S^3$.

\section{Some twisted conjugacy classes in $\SU(2) \times \SU(2)$}

The primary goal of this section is to determine the twisted conjugacy
classes of $\SU(2)\times\SU(2)$ and understand their geometry and
topology.  However we think it might be useful to consider in the
beginning a slightly more general problem, which basically consists in
replacing $\SU(2)$ with an arbitrary group.

Let $G$ be a Lie group and let $D = G \times G$ be the product group.
Let $\tau: D \to D$ be the twist $\tau(x,y) = (y,x)$.  It is clearly
an outer automorphism of $D$.

Let $\Ad_\tau$ denote the twisted adjoint action of $D$ on $D$,
\begin{equation*}
  \Ad_\tau(x,y)\cdot (x_0,y_0) = (x x_0 y^{-1}, y y_0 x^{-1})~,
\end{equation*}
for all $(x,y), (x_0,y_0) \in D$.  By the twisted conjugacy class of a
point $(x,y)\in D$, we mean the orbit of $(x,y)$ under the twisted
adjoint action of $D$ on $D$.  We denote it $\eO_{(x,y)}$.  In other
words,
\begin{equation*}
    \eO_{(x,y)} = \{ (w x z^{-1}, z y w^{-1}) \mid w,z\in G\}~.
  \end{equation*}
In the notation of the last section, $\eO_{(x,y)} = \eC^{\tilde
T}_{G\times G}(x,y)$.

We would like to determine the twisted conjugacy classes of $D$ and
explore their topology.  If, as in the case of interest, $G$ (and
hence $D$) possesses a bi-invariant metric, we also would like to say
something about the geometry of the twisted conjugacy classes as
submanifolds of $D$.

We start with an example.  The twisted conjugacy class of the
identity $(e,e)$ is the ``anti-diagonal'', a submanifold of $D$
diffeomorphic to $G$, but which is not a subgroup:
\begin{equation*}
  \Ad_\tau(x,y)\cdot (e,e) = (x y^{-1}, y x^{-1}) = (z,z^{-1})
\end{equation*}
for $z:=x y^{-1}$.  Hence the orbit $\eO_{(e,e)}$ of $(e,e)$ is
\begin{equation*}
  \eO_{(e,e)} = \{ (z,z^{-1}) | z \in G \} \cong G~,
\end{equation*}
where the isomorphism is one of differentiable manifolds.

This example points the way to determining the rest of the twisted
conjugacy classes.  We start with a series of observations.

Group multiplication gives a natural surjection $m: D = G \times G \to
G$.  We will write it simply as $m(x,y) = xy$.  It is easy to see that
the inverse image of a point is diffeomorphic to $G$.  Indeed, suppose
$xy = x'y'$.  Then $x^{-1} x' = y (y')^{-1} = z$, say.  Therefore, $x'
= x z$ and $y' = z^{-1} y$, for some $z \in G$.  In other words,
\begin{equation*}
  m^{-1} (xy) = \{ (x z, z^{-1} y) \mid z \in G \} \cong G~.
\end{equation*}
This means that $G\times G$ is a bundle over $G$ with fibre $G$, but
the fibration is different than the standard fibrations $\pr_1: G
\times G \to G$ sending $(x,y)\mapsto x$ and $\pr_2: G \times G \to G$
sending $(x,y)\mapsto y$.  Nevertheless $m: G \times G \to G$ is a
principal $G$-bundle.  To prove this we must simply exhibit a free
action of $G$ which preserves the fibres.  The typical fibre is given
by
\begin{equation*}
  m^{-1} (xy) = \{ (x z, z^{-1} y) \mid z \in G \}~.
\end{equation*}
Let $g\in G$ and consider the action $(x,y) \mapsto (xg^{-1},gy)$.
This is clearly free and moreover $x g^{-1} g y = xy$, whence it
preserves the fibre.

We can now state the main result of this section.

\begin{thm}
  Let $\eO_{(x,y)}$ be the twisted conjugacy class of $(x,y) \in D$
  and let $\eC_z$ be the (standard) conjugacy class of $z \in G$.
  Then,
  \begin{equation*}
    \eO_{(x,y)} = m^{-1} \eC_{xy}~.
  \end{equation*}
  In other words, twisted conjugacy classes in $D$ are the inverse
  images under the group multiplication of the (standard) conjugacy
  classes in $G$.
\end{thm}

\begin{proof}
  A typical element in $\eO_{(x,y)}$ is $(uxv^{-1}, vyu^{-1})$.  The
  product of these two elements is $uxyu^{-1}$, which is conjugate to
  $xy\in G$.  In other words, $m(\eO_{(x,y)}) = \eC_{xy}$, whence
  $\eO_{(x,y)} \subseteq m^{-1} \eC_{xy}$.  To prove the reverse
  inclusion, we will prove that if $ab$ and $cd$ are conjugate in $G$,
  then $(a,b)$ and $(c,d)$ belong to the same twisted conjugacy class
  in $D$.  If $ab$ and $cd$ are conjugate, then $cd = zabz^{-1}$ for
  some $z \in G$.  This is equivalent to
  \begin{align*}
    z^{-1}cd = abz^{-1} &\iff a^{-1} z^{-1} c = b z^{-1} d^{-1} =
    w^{-1}\quad\exists w \in G\\
    &\iff c = z a w^{-1}\quad\text{and}\quad d = w b z^{-1}\\
    &\iff (c,d) = (z a w^{-1}, w b z^{-1})~,    
  \end{align*}
  whence $(c,d)$ and $(a,b)$ are in the same twisted conjugacy class.
\end{proof}

As a corollary we have that twisted conjugacy classes in $D=G\times G$
are principal $G$-bundles over conjugacy classes of $G$.

We now specialise to $G = \SU(2)$.  The Lie group $\SU(2)$ has two
kinds of conjugacy classes: points (at $\pm e$) and $2$-spheres
everywhere else.  Now, the inverse image under $m$ of the points are
topologically 3-spheres, whereas the inverse image of a $2$-sphere is
a principal $\SU(2)$-bundle over $S^2$.  Since principal $G$-bundles
over $S^2$ are classified up to homotopy by $\pi_1(G)$, the fact that
$\SU(2)$ is simply-connected implies that the bundle is trivial.  In
other words, the Lie group $\SU(2)\times\SU(2)$ has two types of
twisted conjugacy classes, diffeomorphic to $S^3$ or to $S^2 \times
S^3$.

Notice furthermore that the $S^3$ orbits are homologically nontrivial.
This is because the maps induced in homology by the canonical
projections $\pr_1$ and $\pr_2$ send the homology classes of the
orbits to the fundamental class of $\SU(2)$, up to orientation.

For example, if $\eO = \eO_{(e,e)}$, notice that
\begin{equation*}
  \pr_1(z,z^{-1}) = z \quad\text{and}\quad
  \pr_2(z,z^{-1}) = z^{-1}~.
\end{equation*}
Therefore $(\pr_1)_*[\eO] = 1 \in \Z \cong H_3(S^3)$ and
$(\pr_2)_*[\eO] = -1 \in \Z \cong H_3(S^3)$.

On the other hand, the five-dimensional classes are homologically
trivial, since $H_5(S^3\times S^3) = 0$.  Nevertheless a similar
argument shows that they are not homotopically trivial.

It remains to see how the twisted conjugacy classes are embedded
geometrically relative to the bi-invariant metric on $G \times G$.  We
will first consider the small orbits (of dimension $\dim G$) obtained
from point-like conjugacy classes in $G$.  As we will show, they are
totally geodesic submanifold of $D$; and, in particular, they are
minimal.

Consider $\eO = \eO_{(e,e)}$.  It is not a subgroup of $D$, but we
will see that it is a subgroup of $D'$, a Lie group which shares the
same underlying manifold as $D$, but whose group multiplication is
different.  Moreover, the bi-invariant metric on $D$ is also
bi-invariant in $D'$.  It will follow that $\eO$ is totally geodesic
as a consequence of the following well-known result (see, for example, 
Exercise~6.6 in \cite{DoCarmo}):

\begin{thm}
  Let $G$ be a Lie group with a bi-invariant metric.  Then any
  subgroup $H$ is a totally geodesic submanifold.
\end{thm}

The group $D'$ is defined as $G \times G^{\opp}$, where $G^{\opp}$,
the opposite group, is the group sharing the same underlying manifold
with $G$ but with the opposite multiplication law:
\begin{align*}
  m^{\opp} :  G^{\opp} \times G^{\opp} &\to G^{\opp}\\
              (x,y) &\mapsto yx~.
\end{align*}
In other words, $m^{\opp} = m \circ \tau$.  Clearly $G^{\opp}$ and $G$ 
are isomorphic as Lie groups: the isomorphism $G \to G^{\opp}$ being
defined by $x \mapsto x^{-1}$.

Under the group multiplication in $D' = G \times G^{\opp}$,
\begin{align*}
        m' : D' \times D' &\to D'\\
             ((x,y),(u,v)) &\mapsto (xu, vy)~,
\end{align*}
it is clear that $\eO$ is a subgroup:
\begin{equation*}
  (x,x^{-1}) \cdot (y, y^{-1}) = (xy, y^{-1} x^{-1}) = (xy,
  (xy)^{-1})~.
\end{equation*}

Since $D'$ has the same underlying manifold as $D$, we have a
$D$-bi-invariant metric on it.  To be able to apply the theorem, it
remains to show that this metric is also $D'$-bi-invariant.  The
bi-invariant metric on $D=G \times G$ is the riemannian product of the
\emph{same} bi-invariant metric on each of the factors: this
guarantees that $\tau$ is an isometry.  Therefore this metric on $G
\times G$ will be bi-invariant under $D'$ if and only if the metric on
$G$ is $G^{\opp}$-bi-invariant.  But this follows trivially from the
following observation.  For $x\in G$ let $L(x)$ and $R(x)$ denote the
left- and right-multiplication by $x$ in $G$, respectively.
Similarly, let $L^{\opp}(x)$ and $R^{\opp}(x)$ denote the similar
operations in $G^{\opp}$.  Then one has
\begin{equation*}
  L(x) = R^{\opp}(x) \qquad\text{and}\qquad
  R(x) = L^{\opp}(x)~.
\end{equation*}
This means that left-invariance under $G$ is equivalent to
right-invariance under $G^{\opp}$ and viceversa.  In particular,
bi-invariance under $G$ is equivalent to bi-invariance under
$G^{\opp}$.  We conclude that since the metric on $G$ is
$G$-bi-invariant, it is also $G^{\opp}$-bi-invariant.
In summary, we have just proven the following:

\begin{thm}
  The twisted conjugacy class $\eO_{(e,e)}$ is totally-geodesic
  relative to the bi-invariant metric on $D$.  In particular, it is
  minimal.
\end{thm}

How about the other small twisted conjugacy classes?  These are the
inverse images by the multiplication $m$ of point-like conjugacy
classes in $G$, hence of elements in the centre of $G$.  Let $z$ be an
element in the centre of $G$.  Then the twisted conjugacy class
$\eO_{(e,z)} := m^{-1}(z)$ is given by
\begin{equation*}
  \eO_{(e,z)} = \{ (x, x^{-1}z) \mid x \in G \}~.
\end{equation*}
This is the translate (both left and right) of $\eO_{(e,e)}$ by the
element $(e,z)$.  Since the metric on $D$ is bi-invariant,
$\eO_{(e,z)}$ is isometric to $\eO_{(e,e)}$ as submanifolds of $D$.
In particular, since $\eO_{(e,e)}$ is totally geodesic, so is
$\eO_{(e,z)}$.  This proves the following:

\begin{thm}
Let $z\in G$ be any element in the centre.  The twisted conjugacy class
$\eO_{(e,z)}$ is totally-geodesic relative to the bi-invariant metric
on $D$.  In particular, it is minimal.
\end{thm}

Specialising to $G=\SU(2)$ we have that the twisted conjugacy classes
$\eO_{(e,e)}$ and $\eO_{(e,-e)}$ in $\SU(2) \times \SU(2)$ are
totally-geodesic three-spheres.

How about the larger orbits?  In this case, it is possible to argue
that it is metrically a fibre product with totally-geodesic fibres
diffeomorphic to $G$.  But the total space of the bundle is certainly
not totally geodesic.  In fact, it is in general not even minimal.

\section{The $N{=}1$ supersymmetric extension}

Let us now consider the $N{=}1$ supersymmetric extension of
the affine Lie algebra $\widehat\gg$, which we will denote by
$\widehat\gg_{N{=}1} = \widehat{\mathfrak{sl}}(2,\R)_{N{=}1} \oplus
\widehat{\mathfrak{su}}(2)_{N{=}1}$, with generators $(\J_a,\Psi_a)$
satisfying
\begin{align}
\J_a (z)\J_b (w) &= \frac{h_{ab}}{(z-w)^2} +
         \frac{{f_{ab}}^c \J_c (w)}{z-w} + \reg~,\label{eq:sc1}\\
\J_a (z)\Psi_b (w) &= \frac{{f_{ab}}^c \Psi_c (w)}{z-w} +
         \reg~,\label{eq:sc12}\\
\Psi_a (z)\Psi_b (w) &= \frac{h_{ab}}{z-w} + \reg~,\label{eq:sc2}
\end{align}
with $h_{ab}$ defined as in \eqref{eq:g}.
The free fields $(\varphi,\lambda)$ on $S^1$ satisfy the standard OPEs
\begin{align}
\d\varphi(z)\d\varphi(w) &= \frac{1}{(z-w)^2} + \reg~,\label{eq:sS1}\\ 
\lambda(z)\lambda(w) &= \frac{1}{z-w} + \reg~.\label{eq:sS2}
\end{align}

Then the generators of the $N{=}1$ SCA will be given by
\begin{align*}
\sT(z) &= \half h^{ab}(\tilde\J_a \tilde\J_b) + \half
          h^{ab}(\d\Psi_a\Psi_b) + \half(\d\varphi\d\varphi) +
          \half(\d\lambda\lambda)\\ 
\sG(z) &= h^{ab}(\tilde\J_a\Psi_b) + (\d\varphi\lambda) 
         -\tfrac{1}{6k^2} f^{abc} (\Psi_a \Psi_b \Psi_c)~,
\end{align*}
where we have introduced the so-called decoupled currents, $\tilde\J_a
\equiv \J_a - \tfrac{1}{2}h^{bd}{f_{ab}}^c (\Psi_c\Psi_d)$, in terms
of which the superconformal generators take a relatively simple form.
The coefficients $h^{ab}$ are the components of $h^{-1}$.  The central
charge of this SCFT is given by
\begin{equation*}
c = \frac{3}{2} + \frac{3(k_1-1)}{k_1} + \frac{3}{2} +
    \frac{3(k_2+1)}{k_2} + \frac{3}{2} + \frac{3(k_3+1)}{k_3} +
    \frac{3}{2}~.
\end{equation*}
In order to have a critical superstring theory the levels must satisfy 
the following relation
\begin{equation*}
\frac{1}{k_1} - \frac{1}{k_2} - \frac{1}{k_3} = 0~,
\end{equation*}
in which case $c=15$.  And, since we have a similar structure for the
antiholomorphic sector as well, we actually have a $(1,1)$ SCFT.

The gluing conditions are given by
\begin{equation}
  \label{eq:sDbc} 
  \J_a(z)- {R^b}_a {\bar\J}_b(\bar z) = 0~,\qquad 
  \Psi_a(z) - {S^b}_a {\bar\Psi}_b(\bar z) = 0~,
\end{equation}
where the coefficients ${R^b}_a$ and ${S^b}_a$ are defined by
$R,S:\gg\to\gg$, with $R(Z_a)=Z_b{R^b}_a$ and $S(Z_a)=Z_b{S^b}_a$, for
any $Z_a$ in $\gg$.  These conditions are to be understood as
supersymmetric generalisations of the gluing conditions written down in
Section~\ref{sec:boundary}; henceforth $R$ is taken to be an
automorphism of $\gg$ which preserves the metric.  At this point we do
not need to impose any specific condition on $S$, since this will be
fixed, as we will see in a moment, by supersymmetry considerations.

The gluing conditions \eqref{eq:sDbc} have to satisfy a similar
consistency requirement as in the bosonic case.  In this context,
consistency means that the holomorphic SCFT is set equal to the
antiholomorphic SCFT up to an automorphism of the $N{=}1$ SCA; in
other words, at the boundary we must have
\begin{equation*}
  \label{eq:sci}
  \sT(z) = \bar\sT(\bar z)\qquad\text{and}\qquad \sG(z) =
  \pm\bar\sG(\bar z)~.
\end{equation*}
These conditions have been written down previously in \cite{SDKS}, in
the context of Kazama--Suzuki models.

The first requirement translates into a number of conditions on the
matrices $R$ and $S$.  Thus, from the quadratic terms in the currents
we obtain that
\begin{equation*}
R^T \eta R = \eta~,\qquad S^T \eta R = \pm\eta~,
\end{equation*}
which immediately implies that
\begin{equation}
S = \pm R~,\label{eq:sr}
\end{equation}
as one would expect from supersymmetry.  Further, from the cubic terms 
in the currents we have that
\begin{equation*}
[S(Z_a),S(Z_b)] = \pm S([Z_a,Z_b])~,\qquad 
[R(Z_a),S(Z_b)] = S([Z_a,Z_b])~,
\end{equation*}
which, together with \eqref{eq:sr}, implies that
\begin{equation}
[R(Z_a),R(Z_b)] = R([Z_a,Z_b])~.
\end{equation}
In other words, the conditions that $R$ must satisfy in order for the
corresponding $\D$-brane configurations to preserve superconformal
invariance match exactly the assumptions already made on $R$.
Furthermore, it follows that these gluing conditions preserve the
infinite-dimensional symmetry of the $N{=}1$ current algebra
\eqref{eq:sc1}-\eqref{eq:sc2}. 

Finally, since we know from the bosonic case that $R$ must take one of
two particular forms \eqref{eq:R}, we obtain that $S$ must have a
similar form
\begin{equation}\label{eq:S}
S_I = \pm \begin{pmatrix}
           R_{11} & 0 & 0\\
           0 & R_{22} & 0\\
           0 & 0 & R_{33}
         \end{pmatrix}~,\qquad
S_{II} = \pm \begin{pmatrix}
             R_{11} & 0 & 0\\
             0 & 0 & R_{23}\\
             0 & R_{32} & 0
         \end{pmatrix}~,
\end{equation}
We therefore conclude that every bosonic configuration that we
determined can be made into an $N{=}1$ supersymmetric configuration
without having to impose additional conditions.

\section{Spacetime supersymmetry}

In this section we analyse the fraction of spacetime supersymmetry
preserved by the $\D$-brane configurations we determined before.  In the
context of superconformal field theories spacetime supersymmetry
appears as a by-product of $N{=}2$ superconformal invariance, being
related, via bosonisation, to the $U(1)$ current.  Instead of
following this standard approach, here we will analyse the spacetime
symmetry preserved by the $\D$-branes we found using a different route,
which was described in \cite{GKS,EFGT}.

We will therefore consider the spacetime supercharges to be
constructed directly from the $N{=}1$ SCFT.  To this end we introduce
the fermionic fields $\psi_i$ for $\SL(2,\R)$, $\chi_i$ and $\omega_i$
for the two copies of $\SU(2)$, with $i=1,2,3$.  Further, we choose
five fermion bilinears and bosonise them into five scalar fields
$H_I$, with $I=1,\ldots,5$ as follows
\begin{equation*}
\d H_1 = \psi_1\psi_2~,\qquad 
\d H_2 = \chi_1\chi_2~,\qquad 
\d H_3 = \omega_1\omega_2~,
\end{equation*}
\begin{equation*}
\d H_4 = i \left(\sqrt{\frac{k_1}{k_2}}\chi_3 +
         \sqrt{\frac{k_1}{k_3}}\omega_3\right)\psi_3~,\qquad  
\d H_5 = \left(\sqrt{\frac{k_1}{k_2}}\chi_3 -
         \sqrt{\frac{k_1}{k_3}}\omega_3\right)\lambda~.
\end{equation*}

The corresponding spacetime supercharges \cite{FMS} will be required
to be BRST invariant and to pass the GSO projection.  This will yield
\begin{equation}
Q = \oint dz e^{-\frac{\phi}{2}} S(z)~,
\end{equation}
where $\phi$ is the scalar field which appears in the bosonised
superghost system of the fermionic string, and the corresponding spin
fields are given by (for a detailed discussion see \cite{EFGT})
\begin{align*}
S_1(z) &= e^{\frac{i}{2}(H_1 + H_2 + H_3 + H_4 + H_5)}~,\\
S_2(z) &= e^{\frac{i}{2}(H_1 - H_2 - H_3 - H_4 - H_5)}~,\\
S_3(z) &= e^{\frac{i}{2}(-H_1 + H_2 + H_3 - H_4 + H_5)}~,\\
S_4(z) &= e^{\frac{i}{2}(-H_1 - H_2 - H_3 + H_4 - H_5)}~,
\end{align*}
\begin{align*}
S_5(z) &= \sqrt{\frac{k_1}{k_2}}e^{\frac{i}{2}(H_1 - H_2 + H_3 - H_4 +
H_5)} + \sqrt{\frac{k_1}{k_3}}e^{\frac{i}{2}(H_1 - H_2 + H_3 + H_4 -
H_5)}~,\\ 
S_6(z) &= \sqrt{\frac{k_1}{k_2}}e^{\frac{i}{2}(H_1 + H_2 - H_3 + H_4 -
H_5)} + \sqrt{\frac{k_1}{k_3}}e^{\frac{i}{2}(H_1 + H_2 - H_3 - H_4 +
H_5)}~,\\ 
S_7(z) &= \sqrt{\frac{k_1}{k_2}}e^{\frac{i}{2}(-H_1 - H_2 + H_3 + H_4 +
H_5)} + \sqrt{\frac{k_1}{k_3}}e^{\frac{i}{2}(-H_1 - H_2 + H_3 - H_4 -
H_5)}~,\\ 
S_8(z) &= \sqrt{\frac{k_1}{k_2}}e^{\frac{i}{2}(-H_1 + H_2 - H_3 - H_4 -
H_5)} + \sqrt{\frac{k_1}{k_3}}e^{\frac{i}{2}(-H_1 + H_2 - H_3 + H_4 +
H_5)}~.
\end{align*}

For any given $\D$-brane configuration, the boundary conditions satisfied
by the fermionic fields will lead to a certain set of boundary
conditions satisfied by the supercharges.  Let us consider the those
configurations described by $R_I$, with $R_{11}$ of the form of a
spatial rotation.  The corresponding boundary conditions satisfied by
the fermions read (we use the notation introduced in \cite{Sads3})
\begin{align*}
\psi_1 - \cos\alpha \bar\psi_1 - \sin\alpha \bar\psi_2 = 0~,\qquad &
\chi_1 - \cos\beta \bar\chi_1 - \sin\beta \bar\chi_2 = 0~,\\
\psi_2 + \sin\alpha \bar\psi_1 - \cos\alpha \bar\psi_2 = 0~,\qquad &
\chi_2 + \sin\beta \bar\chi_1 - \cos\beta \bar\chi_2 = 0~,\\
\psi_3 - \bar\psi_3 = 0~,\qquad & \chi_3 - \bar\chi_3 = 0~,\\
\omega_1 - \cos\gamma \bar\omega_1 - \sin\gamma \bar\omega_2 =
0~,\qquad & \\
\omega_2 + \sin\gamma \bar\omega_1 - \cos\gamma \bar\omega_2 =
0~,\qquad & \lambda \pm \bar\lambda = 0~,\\
\omega_3 - \bar\omega_3 = 0~,\qquad &
\end{align*}
where we have systematically ignored a $\pm$ sign coming from
\eqref{eq:S}, which does not affect the fermion bilinears in the
expression of $S(z)$.  The $\pm$ sign in the boundary condition on
$\lambda$ corresponds to having Neumann or Dirichlet boundary
conditions on $S^1$, respectively.  We will see however that only one
choice will give rise to states preserving some spacetime
supersymmetry.  From the above relations it follows that the scalar
fields $H_I$ satisfy the following boundary conditions:
\begin{equation*}
H_I = \bar H_I~,\quad I=1,2,3,4~,\qquad H_5=\pm\bar H_5~,
\end{equation*}
where the $\pm$ sign in the boundary condition for $H_5$ reflects the
one in the boundary condition along $S^1$.  Therefore, in order to
preserve some spacetime supersymmetry, we have to impose a Dirichlet
boundary condition along the flat direction of the target space.  In
this way we obtain that these configurations describe odd-dimensional
$\D$-branes (as we expect in the case of a type IIB theory).  It
immediately follows that, for these configurations, the spacetime
supercharges satisfy
\begin{equation}\label{eq:BPS1}
Q_{\alpha} = \bar Q_{\alpha}~,\qquad\qquad \alpha=1,\ldots,8~.
\end{equation}
Hence we conclude that the $\D$-brane configurations defined by $R_I$,
with $R_{11}$ a spatial rotation in $\SO(2,1)$, and characterised by a
Dirichlet condition along $S^1$ preserve half of the spacetime
supersymmetry, and therefore are BPS states.

In order to analyse the configurations described by $R_I$ with
$R_{11}$ given by a boost in $\SO(2,1)$ we need a slight change in the 
way we define the fermion bilinears and the corresponding
supercharges.  Essentially, we need to switch the places of $\psi_2$
and $\psi_3$ in the definition of $H_1$ and $H_4$.  Then by using the
following boundary conditions for the fermions on $\AdS_3$
\begin{align*}
\psi_1 - \cosh\alpha \bar\psi_1 - \sinh\alpha \bar\psi_3 = 0~,\\
\psi_2 - \bar\psi_2 = 0~,\\
\psi_3 - \sinh\alpha \bar\psi_1 - \cosh\alpha \bar\psi_3 = 0~,
\end{align*}
and unchanged conditions for all the other fermions, we obtain that
$H_I = \bar H_I$, for all $I$, provided we set again a Dirichlet
boundary condition along $S^1$.  From this it follows that the
spacetime supercharges satisfy \eqref{eq:BPS1} as before.  Hence all
the corresponding odd-dimensional $\D$-branes describe BPS states that
preserve half of the spacetime supersymmetry.

Finally, in the case where $R_{11}$ is given by a null rotation in
$\SO(2,1)$, due to the particular form of the boundary conditions and
to the nonlocal nature of the dependence of the spacetime supercharges
on the fermionic fields, it is rather difficult to determine the
fraction of spacetime supersymmetry preserved by this particular type
of boundary states.

Let us now turn to the $\D$-brane configurations described by $R_{II}$.
If we start with configurations characterised by a component $R_{11}$
of the form of a spatial rotation in $\SO(2,1)$ then the corresponding
boundary conditions on the fermions read
\begin{align*}
\psi_1 - \cos\alpha \bar\psi_1 - \sin\alpha \bar\psi_2 = 0~,\qquad &
\chi_1 - \cos\beta \bar\omega_1 - \sin\beta \bar\omega_2 = 0~,\\
\psi_2 + \sin\alpha \bar\psi_1 - \cos\alpha \bar\psi_2 = 0~,\qquad &
\chi_2 + \sin\beta \bar\omega_1 - \cos\beta \bar\omega_2 = 0~,\\
\psi_3 - \bar\psi_3 = 0~,\qquad & \chi_3 - \bar\omega_3 = 0~,\\
\omega_1 - \cos\gamma \bar\chi_1 - \sin\gamma \bar\chi_2 = 0~,\qquad &
\\
\omega_2 + \sin\gamma \bar\chi_1 - \cos\gamma \bar\chi_2 =
0~,\qquad & \lambda \pm \bar\lambda = 0~,\\
\omega_3 - \bar\chi_3 = 0~,\qquad &
\end{align*}
This time, in order to be able to preserve some fraction of the
spacetime supersymmetry, we must impose a Neumann boundary condition 
along $S^1$.  Then we obtain
\begin{equation*}
H_1 = \bar H_1~,\quad H_2 = \bar H_3~,\quad H_3 = \bar H_2~,
\quad H_4 = \bar H_4~,\quad H_5 = \bar H_5~.
\end{equation*}
This in turn implies that the corresponding supercharges (where, we
recall, $k_2=k_3$) will satisfy the following conditions
\begin{equation}\label{eq:BPS2}
Q_{\alpha} = {A^{\beta}}_{\alpha}\bar Q_{\beta}~,\qquad
                                        \alpha,\beta=1,\ldots,8 
\end{equation}
where the coefficients ${A^{\beta}}_{\alpha}$ are the elements of the
matrix $A$
\begin{equation*}
A = \begin{pmatrix}
            \1 & & & \\
            & \1 & & \\            
            & & \sigma & \\            
            & & & \sigma            
         \end{pmatrix}~,
\end{equation*}
where $\sigma = \begin{pmatrix} 0 & 1\\ 1 & 0 \end{pmatrix}$ and
$\1$ is the two-dimensional identity matrix.  This means that these
configurations too preserve half the spacetime supersymmetry and thus
constitute BPS states.

Similar results hold in the case where the component $R_{11}$ of
$R_{II}$ is given by a boost, provided we use the appropriate choice
for the scalar fields $H_I$.  And also similarly, we can not say much
about the boundary states having the $R_{11}$ of the form of a null
rotation.

The results are summarised in Table~\ref{tab:Qbc}.
\begin{table}[h!]
\scriptsize
\renewcommand{\arraystretch}{1.4}
\let\p=\phantom
\begin{tabular}{|>{$}c<{$}|>{$}c<{$}|>{$}l<{$}|>{$}l<{$}|}
\hline
\text{D7} & (\R\times S^1)\times S^2\times S^3\times S^1 & H_I=\bar
H_I,\ I=1,4,5 & \quad Q_{\alpha} = {A^{\beta}}_{\alpha}\bar Q_{\beta}\\  
 & \R^2\times S^2\times S^3\times S^1 & H_2=\bar H_3,\ H_3=\bar H_2 & 
\quad \alpha,\beta=1,...,8\\
\hline
\text{D5} & (\R\times S^1)\times S^2\times S^2 & H_I=\bar
H_I,\ I=1,...,5 & Q_{\alpha} = \bar Q_{\alpha},\ \alpha=1,...,8\\ 
 & \R^2\times S^2\times S^2 & & \\ 
\hline
 & \{\pm e\}\times S^2\times S^3\times S^1 & & \\
\text{D5} & (\R\times S^1)\times\{\pm e\}\times S^3\times S^1 & H_I=\bar
H_I,\ I=1,4,5 & \quad Q_{\alpha} = {A^{\beta}}_{\alpha}\bar Q_{\beta},\\ 
 & \R^2\times\{\pm e\}\times S^3\times S^1 & H_2=\bar H_3,\ H_3=\bar
 H_2 &\quad \alpha,\beta=1,...,8 \\
\hline
 & \{\pm e\}\times S^2\times S^2 & & \\
\text{D3} & (\R\times S^1)\times\{\pm e\}\times S^2 & H_I=\bar H_I,\
I=1,...,5 & Q_{\alpha} = \bar Q_{\alpha},\ \alpha=1,...,8\\ 
 & \R^2\times\{\pm e\}\times S^2 & & \\ 
\hline
\text{D3} & \{\pm e\}\times\{\pm e\}\times S^3\times S^1 & H_I=\bar
H_I,\ I=1,4,5 & \quad Q_{\alpha} = {A^{\beta}}_{\alpha}\bar Q_{\beta},\\ 
 &  & H_2=\bar H_3,\ H_3=\bar H_2 &\quad \alpha,\beta=1,...,8 \\
\hline
 & \{\pm e\}\times\{\pm e\}\times S^2 & &\\
\text{D1} & (\R\times S^1)\times\{\pm e\}\times\{\pm e\} & H_I=\bar
H_I,\ I=1,...,5 & Q_{\alpha} = \bar Q_{\alpha},\ \alpha=1,...,8\\
 & \R^2\times\{\pm e\}\times\{\pm e\} & & \\
\hline
\text{D(-1)} & \{\pm e\}\times\{\pm e\}\times\{\pm e\} & H_I=\bar
H_I,\ I=1,...,5 & Q_{\alpha} = \bar Q_{\alpha},\ \alpha=1,...,8\\
\hline
\end{tabular}
\vspace{8pt}
\caption{Spacetime supersymmetry for $\D$-brane
configurations in $\AdS_3\times S^3\times S^3\times S^1$.\label{tab:Qbc}}
\end{table}

\section{Conclusions}

In this paper we have studied, using the SCFT framework and the
boundary state formalism, the possible $\D$-brane configurations which
one can consistently define in an $\AdS_3\times S^3\times S^3\times
S^1$ background characterised by a purely NS-NS B field.  

We have analysed a certain type of gluing conditions (type-D,
according to the nomenclature used in \cite{SDnotes,Sads3}), which are
characterised by the fact that they preserve not only the
superconformal structure of the background but also the underlying
symmetry of the $N{=}1$ current algebra.  We have seen that the
solutions fall in two different classes.  The first class, produced by
gluing conditions defined in terms of inner automorphisms of the
corresponding Lie algebra, describes D-brane configurations whose
worldvolumes are products of shifted conjugacy classes
\begin{equation*}
\eC_{\SL(2,\R)}r_1 \times \eC_{\SU(2)}r_2\times \eC_{\SU(2)}r_3~,
\end{equation*}
giving thus rise to odd-dimensional D-branes embedded in
$\SL(2,\R)\times\SU(2)\times\SU(2)$ and even-dimensional D-branes
wrapped on the flat $S^1$.  It is however the odd-dimensional D-branes
that turn out to also preserve half of the spacetime supersymmetry of
the background.

The second class of solutions, produced by gluing conditions defined
in terms of outer automorphisms, describes D-brane configurations
whose worldvolumes are products of shifted conjugacy classes in
$\SL(2,\R)$ with twisted conjugacy classes in
$\times\SU(2)\times\SU(2)$
\begin{equation*}
\eC_{\SL(2,\R)}r_1 \times \eC^{\tilde T}_{\SU(2)\times\SU(2)}\tilde r~.
\end{equation*}
We have studied in some detail the twisted conjugacy classes of
$\SU(2)\times\SU(2)$, showing that they can be characterised as the
inverse images, under the group multiplication, of the standard
conjugacy classes of $\SU(2)$.  In particular, they consist of two
totally-geodesic three-spheres and a family of homologically trivial
but homotopically nontrivial five-dimensional submanifolds
diffeomorphic to $S^2\times S^3$.  These solutions give rise to
even-dimensional branes embedded in
$\SL(2,\R)\times\SU(2)\times\SU(2)$ and odd-dimensional D-branes
wrapped on the flat $S^1$, and it is again the odd-dimensional branes
that preserve half of the spacetime supersymmetry.

\section*{Acknowledgements}

It is a pleasure to thank AA~Tseytlin for many useful discussions in
the early stages of this project.  This work was partially supported
by a PPARC Postdoctoral Fellowship.

%
%
\providecommand{\href}[2]{#2}\begingroup\raggedright\endgroup

\end{document}